\documentclass[twoside]{llncs}

\usepackage{graphicx} 
\usepackage{amsmath}  
\usepackage{array} 
\usepackage[colorlinks=true,%
            linkcolor=blue,%
            citecolor=blue,%
            urlcolor=blue]{hyperref}           

\usepackage{fancyhdr}           
\usepackage{ifthen}

\newcommand\settitle[2][]{%
 \title{#2}
 \ifthenelse{\equal{#1}{}}%
  {\fancyhead[RO]{\nouppercase #2 \qquad \thepage}}%
  {\fancyhead[RO]{\nouppercase #1 \qquad \thepage}}%
}

\newcommand\setauthors[2]{%
 \author{#2}
  {\fancyhead[LE]{\thepage \qquad \nouppercase #1}}%
}

\fancypagestyle{plain}{%
\fancyhf{}  

}

\def\keywordsname{Keywords.}
\newenvironment{keywords}{%
      \list{}{\advance\topsep by-0.50cm\relax\small
      \leftmargin=1cm
      \labelwidth=1cm
      \listparindent=1cm
      \itemindent\listparindent
      \rightmargin\leftmargin}\item[\hskip\labelsep
                                    \bfseries\keywordsname]}
    {\endlist}

\begin{document}


\settitle
         {Spanning Trees of Bounded Degree Graphs}


\setauthors{Robson}
           {John Michael Robson\inst{1}
           }
     
\institute{LaBRI, Universit\'e Bordeaux 1\\
  351 cours de la Lib\'eration, 33405 Talence Cedex, France\\
\email{robson@labri.fr}
}

\date{}
\maketitle

\thispagestyle{plain}

\begin{abstract}
We consider lower bounds on the number of spanning trees of connected graphs
with degree bounded by $d$.
The question is of interest because such bounds may improve the analysis of the
improvement produced by memorisation in the runtime of exponential algorithms.
The value of interest is the constant $\beta_d$ such that all connected graphs with degree bound
ed
by $d$ have at least $\beta_d^\mu$ spanning trees where $\mu$ is the cyclomatic number or
excess of the graph, namely $m-n+1$.

We conjecture that $\beta_d$ is achieved by the complete graph $K_{d+1}$ but we have
not proved this for any $d$ greater than $3$. We give weaker lower bounds on $\beta_d$
for $d\le 11$.

First we establish lower bounds on the factor by which the number of spanning trees
is multiplied when one new vertex is added to an existing graph so that the new vertex
has degree $c$ and the maximum degree of the resulting graph is at most $d$. In all
the cases analysed, this lower bound $f_{c,d}$ is attained when the graph before the addition wa
s
a complete graph of order $d$ but we have not proved this in general. 

Next we show that, for any cut of size $c$ cutting a graph $G$ of degree bounded by $d$
into two connected components $G_1$ and $G_2$, the number of spanning trees of $G$ is
at least the product of this number for $G_1$ and $G_2$ multiplied by the same
factor $f_{c,d}$.

Finally we examine the process of repeatedly cutting a graph until no edges remain.
The number of spanning trees is at least the product of the multipliers associated with
all the cuts. Some obvious constraints on the number of cuts of each size give
linear constraints on the normalised numbers of cuts of each size which are then used
to lower bound $\beta_d$ by the solution of a linear program.
The lower bound obtained is significantly improved by
imposing a rule that, at each stage, a cut of the minimum available size is chosen
and adding some new constraints implied by this rule.

\end{abstract}


\section{Introduction \label{intro}}
We consider lower bounds on $SP(G)$ the number of spanning trees of a connected graph $G$.

Clearly a tree has only one spanning tree and adding a single edge to a tree
creates a cycle which can be broken in at least $3$ ways giving $3$ spanning trees.
Adding a second edge does not necessarily multiply $SP(G)$ by $3$ again
since a square with one diagonal edge has $8$ spanning trees rather than $9$.

We are interested in lower bounds which are exponential in the number of edges added,
that is the cyclomatic number of the graph,
but no such bound can exist for general graphs. Accordingly we consider
graphs for which an upper bound holds on the maximum degree.

This study was motivated by the analysis of the effectiveness of memorisation
in reducing the computation time of some graph algorithms, effectiveness which
depends on the number of small induced subgraphs encountered (\cite{robson}).
The most effective way known to upper bound this number of small induced subgraphs 
is to count the number of their spanning trees; knowing that each
subgraph has many spanning trees enables us to reduce the upper bounds so obtained.

We will make considerable use of two well known properties of a spanning tree chosen
uniformly
\begin{itemize}
\item The {\it electrical property:} the probability that an edge $(u,v)$ is included
in the spanning tree is $1/(1+res(u,v))$ where $res(u,v)$ is the resistance
between $u$ and $v$ of an electrical
network obtained by deleting the edge $(u,v)$ and replacing every other edge by a
$1$ ohm resistor,
\item The {\it random walk model:} the tree is exactly that produced by a random walk
on the graph where an edge traversed in the random walk is added to the tree precisely
if it arrives at a node not already in the tree.
\end{itemize}


\section{Some definitions and a conjecture}
We define the excess edges of a connected graph $G=(V,E)$ as the number
of edges minus the number in a spanning tree, that is the cyclomatic number: $\mu (G)=|E|-|V|+1$
\\
Then $\beta(G)~=~SP(G)^{1/\mu (G)}$ is the geometric mean of the factors by which $SP(G)$ is
multiplied in adding the excess edges.\\
Then we define $\beta_d$ as the minimum of $\beta(G)$ over all graphs $G$ with vertex
degrees at most $d$.\\
We conjecture (Conjecture 1) that $\beta_d$ is attained by $K_{d+1}$.
$SP(K_{d+1})=(d+1)^{d-1}$ and $\mu (K_{d+1})=d(d-1)/2$ so that our
conjecture is that $\beta_d~=~(d+1)^{2/d}$.
The fact that $\beta(K_{d+1})~=~(d+1)^{2/d}$ justifies the remark in the Introduction
that no lower bound  ($>1$) holds for $\beta(G)$ in general.\\
We will show lower bounds on $\beta_d$ for $d\le 11$ which are somewhat weaker than this
conjecture.

\section{Lower bounds}
\subsection{A General Lower Bound}
Since adding a new vertex of degree $d$ mutiplies $SP(G)$ by at least $d$ and increases $\mu(G)$
by exactly $d-1$, we have a simple lower bound of $d^{1/(d-1)}$ for $\beta_d$ which is
obviously rather weak because a graph of maximum degree $d$ cannot be built up
by repeatedly adding new vertices of degree $d$. This section
will strengthen this bound for small $d$.

\subsection{Adding a vertex}

We first consider the effect on $SP(G)$ of adding a new vertex.

When a new vertex $v$ of degree $c$ is added, the number of spanning trees is
obviously multiplied
by at least $c$. The multiplying factor is in fact
lower bounded by $f_{c,d}$ strictly greater than $c$, given an upper
bound $d$ on the degree of the graph (after the addition).\\
Conjecture 2: $f_{c,d}$ is achieved when $G$ is $K_d$.

Consider a graph $G$ with $c$ distinguished vertices $u_i~1\le i\le c$
and $G^\prime$ consisting of $G$, a new vertex $v$ and $c$ new edges $(v,u_i)$.
Define the multiplying factor $f (G) = SP (G ^ \prime) / SP (G) $.
When $G$ is $K_d$, we can prove by induction, using the electrical property,
that $SP(G^\prime)=cd^{d-1-c}(d+1)^{c-1}$
so that our conjecture is that $f_{c,d}~= c((d+1)/d)^{c-1}$.

Lemma: Conjecture 2 is true for $c\le d\le 11$.

Proof: 
First we claim that $f(G)$ is decreased by adding any new edge to $G$.
This can be deduced from the electrical property or it is a consequence of the
more general result of \cite{FeMi} Lemma 3.2
which shows that the event $e \in T$ ($T$ a random spanning tree) is
negatively associated with any monotone combination of other such events.
Therefore adding $e$ makes $v$ more likely
to be a leaf and so decreases the ratio
$SP(G^\prime)/SP(G^\prime\setminus v)=SP(G^\prime)/SP(G)=f(G)$.

Also $f(G)$ is not changed by adding a new vertex to $G$ connected to one existing vertex,
so adding a new vertex connected to two or more existing vertices again decreases $f(G)$.

Defining $G_k$ for any $k\ge |G|-c$ as $G$ with new vertices and edges added
so that it consists
of the $u_i$ still with their same induced subgraph
together with a $k$-clique 
and enough edges into the clique from each $u_i$ to make its degree $d-1$, we conclude that
$f(G)\ge f(G_k) \ge f(G_{k+1})$. Then we consider the limit as $k\rightarrow \infty$. 
Considering the random walk model  of a random spanning tree,
we see that in the limit $G_k$ behaves exactly like a weighted graph $W$ consisting of all the $
u_i$
with their same induced subgraph
and a single vertex $w$ connected to each $u_i$ by an edge of weight ($d-1$ minus the degree of 
$u_i$
in this induced subgraph). Thus $f(G) \ge f(W)$ where $W$ depends only on
the subgraph of $G$ induced on $\{u_i\}$.

Now a lower bound on $f(G)$ can be computed by a
(lengthy) computation over all possible induced subgraphs of $c$ vertices with degree
less than $d$. For $c$ up to $10$, the possible subgraphs were generated by a relatively
simple program. For the $1018997864$ cases when $c=11$ we used Brendan McKay's $geng$
program (\cite{bdm}).

The results of this computation are shown in the table. In each case the smallest value of $f$
was given by the induced subgraph $K_{c}$ so that the lower bound is strict,
being given, by another application of the random walk argument, by any graph $G$ in which
the vertices of the $K_{c}$ are all connected to the same $d-c$ other vertices whatever
the edges between these other vertices,
and in particular by $G~=~K_d$, in accordance with the conjecture.
\begin{table}
\begin{tabular}{r|ccccc}
$c$&2&3&4&5&6\\ \hline
d=3&2.666667&5.333333\\
     4&2.500000&4.687500&7.812500\\
     5&2.400000&4.320000&6.912000&10.368000\\
     6&2.333333&4.083333&6.351852&9.263117&12.968364\\
     7&2.285714&3.918367&5.970845&8.529779&11.697983\\
     8&2.250000&3.796875&5.695313&8.009033&10.812195\\
     9&2.222222&3.703704&5.486968&7.620790&10.161053\\
     10&2.200000&3.630000&5.324000&7.320500&9.663060\\
     11&2.181818&3.570248&5.193088&7.081484&9.270306\\ \hline \hline
\end{tabular}
\vspace{3mm}
\begin{tabular}{r|ccccc}
$c$&7&8&9&10&11\\ \hline
   d=7&15.597311\\
     8&14.191006&18.24557\\
     9&13.171735&16.726013&20.907516\\
     10&12.400927&15.589737&19.292299&23.579477\\
     11&11.798571&14.709907&18.053067&21.882506&26.259007\\ \hline
\end{tabular}
\caption{The multiplying factors $f_{c,d}$}
\end{table}

\subsection{Cuts}

Lemma: 
If a graph $G$ (of maximum degree $\le d \le 11$) is cut into two components $G_1$ and $G_2$ by 
the
removal of $c$ edges ($c<d$), $SP(G)\ge f_{c,d}SP(G_1)SP(G_2)$.

{\it Note} In fact this result is also true for $c=d$ but the proof given below does
not cover this case for all $d\le 11$ and we prefer not to give a more complex proof
when we need the result only for $c<d$. 
Proof:
(We write $\Pi$ for $SP(G_1)*SP(G_2)$).
Let the endpoints of the cut edges in $G_1$ and $G_2$ be $U$ and $V$ respectively. 
If all the endpoints of the cut edges in one of the components (say $G_1$) are distinct,
a different spanning tree of $G$ is given by the union of the edges of any spanning tree
of $G_1+v$ and any spanning tree of $G_2$ where, by $G_1+v$ we mean the graph consisting
of $G_1$ together with a vertex $v$ connected to each vertex of $U$. So, in this case,
the lemma is true for all $d$.

In the particular case of two edges $(u_1,~v_1)$ and $(u_2,~v_2)$ with $u_1\not=u_2$ and
$v_1\not=v_2$, we can do better: there are at least $f_{2,d}\Pi$ spanning trees containing a
spanning tree of $G_1$ and at least this same number containing a spanning tree of $G_2$.
Of these exactly $2\Pi$ occur in both the sets (those consisting of a spanning
tree of each component plus one of the edges $(u_i,~v_i)$) so that there are at least
$(2f_{2,d}-2)\Pi$ spanning trees of $G$.

In the general case we consider the bipartite graph $C$ of the cut consisting of $c$ edges
joining  $U$ and  $V$. Without loss of generality
we suppose that $u_1$ has the highest degree $maxu$ (in $C$) of all vertices of $U$,
that $v_1$ has the highest degree $maxv$ (in $C$) of all vertices of $V$ and that
$maxu\ge maxv$. We give a lower bound on the number of spanning trees of $G$
having one of the following
forms:
\begin{itemize}
\item trees with exactly one cut edge ($ c\Pi$)
\item trees with at least 2 cut edges with a common end point at $u_1$ or $v_1$
(and no other cut edges)
($ (f_{maxu,d}-maxu)\Pi$ and $ (f_{maxv,d}-maxv)\Pi$)
\item for every remaining pair of cut edges, trees containing exactly
that pair ($ (f_{2,d}-2)\Pi$ if the pair have a common end point and $ 2(f_{2,d}-2)\Pi$ otherwis
e)
\end{itemize}
The number of pairs of edges with a common end point other than $u_1$ or $v_1$
is $\sum_{i=2}^{|U|}\binom{d_C(u_i)}{2}+\sum_{i=2}^{|V |}\binom{d_C(v_i)}{2}$.
For given $maxu$ and $maxv$, our lower bound is thus
$(c+f_{maxu,d}-maxu+f_{maxv,d}-maxv + (f_{2,d}-2)(2(\binom{c}{2}-\binom{maxu}{2}-\binom{maxv}{2}
)-
\sum_{i=2}^{|U|}\binom{d_C(u_i)}{2}-\sum_{i=2}^{|V |}\binom{d_C(v_i)}{2})\Pi$

This expression is minimised when the degrees $d_C(u_i)$ and $d_C(v_i)$ are chosen according to 
the ``greedy''
partition, that is (for instance for $U$) the lexicographically greatest partition of $c$ into
positive parts respecting the necessary constraints $d_C(u_i)\le d_C(u_1)=maxu$ and $|U|\ge maxv
$.
To verify that the lower bound obtained is always at least $f_{c,d}$ it suffices to test that it
 is so
for every combination $2\le c < d\le 11,~ maxu\ge maxv\ge 2,~maxu+maxv\le c+1$
for their respective greedy partitions.
The $200$ relevant conditions along with their greedy partitions are given in an appendix.

\subsection{Dissecting a graph}

We consider the process of cutting a graph
of maximum degree $d$ until nothing remains but singleton vertices.\\
Using the previous result, the number of spanning trees of the
original graph is at least the product of the multipliers associated with each cut.

At each cut we choose one of the available cuts of minimum size. As a result,
the initial cut has size at most $d$
(which can only happen if the graph is $d$-regular) and
all subsequent cuts have size at most $d-1$.

For each possible size $c$ of cut we note its impact on the number of components
(increased by $1$),
the number of edges (reduced by $c$) and the product of the multipliers
(multiplied by $f_{c,d}$).

\subsection{Linear Programming}

We write the obvious constraints on the number of cuts $n_{c}$ of each size $c$, 
that the total number of cuts is $n-1$, $\sum_{c=1}^dn_{c}~= ~n-1$
and the total number of edges cut is $m$, $\sum_{c=1}^dn_{c}c~= ~m$.
We deduce that $\sum_{c=1}^dn_{c}(c-1)~=~\mu (G)$.
We have also the constraints that $n_d~\le~1$ and that $n_d~=~0$ if the graph is not
$d$-regular.

Then we divide
by the excess to give constraints on $x_{c}$ the normalised number of cuts of each size
and use linear programming to solve for (a lower bound on) the logarithm of the
product of multipliers obtained under the constraint $\sum_{c=1}^dx_{c}(c-1)~=~1$.
The constraints on $n_d$ give us that $x_d~\le~1/min$ where $min$ is the minimum excess
of any $d$-regular graph, from which we exclude $K_{d+1}$ for which the conclusion is
already known to be true. For instance for $d=10$, $min=49$ given by the $10$-regular
graph of order $12$.

\subsubsection{Regular graphs}

The critical case is that of certain $d$-regular graphs, namely those for which
the first cut is a $d$-cut, and we first look in detail at this case.
In this case, from the constraint $\sum_{c=1}^dn_{c}~= ~n-1$, we obtain
$\sum_{c=1}^dx_{c} \ge (n-1)/(dn/2-n+1)$ for the smallest $n$
such that a $d$-regular $n$ vertex graph
exists (other than $K_{d+1}$), namely $d+2$ for even $d$ and $d+3$ for odd $d$.
For $d=10$ this gives us $10/49$.

The solution to the linear program would give us a lower bound on $\log(\beta_d)$
if it was also valid for the remaining graphs (those with an initial cut less than $d$).
For instance for $d=10$, the solution is $0.366508$
(giving $\beta_{10}\ge 1.44269$) with a mixture of 1-cuts, 9-cuts and 10-cuts but no others.
We improve on this by noting that such a combination of cuts cannot arise
with our rule of always taking the smallest cut available. For this we need a lemma:

Lemma (the average cut lemma):\\
The average size of all cuts of size less than $k$ after some $k$-cut
($k\le d$) is at least $k/2$\\
Proof: Consider any $j$-cut ($j\ge k$); it splits some connected subgraph
into $2$ components $C_1$ and $C_2$ and all other connected subgraphs are
$(j-1)$-connected. In any following sequence of $c$ cuts not including a
$(\ge j)$-cut, all cuts are within $C_1$ or $C_2$ and so they are split
into $c+2$ components.
Before the preceding cut, each of these components had at least $j$
outgoing edges (otherwise there would have been a $(j-1)$-cut available);
this gives at least $j(c+2)/2$ edges of which
$j$ were removed by the preceding cut. Hence $jc/2$ edges must have been removed by
the sequence of $c$ cuts;  hence at least $j/2$ edges
are removed on average by each cut; hence average cut size $\ge k/2$.

With this added constraint we get a significantly better bound on $\log \beta_d$.
Table 2 gives the lower bounds on $\beta_d$ so obtained and, for comparison,
the upper bounds given by $K_{d+1}$.

For example for $d=10$ this gives the linear program\\
Minimise $0.788457x_2   +1.289233x_3 +1.672225x_4 +1.990679x_5 +2.268310x_6 
+2.517771x_7 +2.746613x_8 +2.959706x_9 +3.160377x_{10} $ under the constraints
\begin{eqnarray*}
x_1-x_2  &  \le  & -0\\
2x_1-2x_3  &  \le  & 0\\
3x_1 -x_3-3x_4 &  \le  & 0\\
4x_1 -2x_4-4x_5& \le  & 0\\
5x_1 +x_3 -x_4-3x_5-5x_6 & \le  & 0\\
6x_1 +2x_3-2x_5-4x_6-6x_7& \le  & 0\\
7x_1 +3x_3+x_4-x_5-3x_6-5x_7-7x_8& \le  & 0\\
8x_1 +4x_3+2x_4-2x_6-4x_7-6x_8-8x_9& \le  & 0\\
x_2 +2x_3+3x_4+4x_5+5x_6+6x_7+7x_8+8x_9+9x_{10}& \ge  & 1\\
49x_{10} & \le & 1
\end{eqnarray*}
(In solving this program we use the values of $\log f_{c,d}$ computed as accurately
as possible rather than these 6 figure approximations.)

A small improvement could be made by the following observation. The last cut other
than $1$-cuts must be a $2$-cut which cuts a cycle into two components. The multiplier
of this cut should thus be $3$ rather than $f_{2,d}$. Writing $x_2^\prime$ for the (normalised)
number of such cuts, we observe that $x_2^\prime \ge x_{10}$ and adjust the objective
function to $\log(3)$ for the new variable. 
In fact for $d>3$
this improvement improves the constant found for regular graphs to  one better
than that for the non-regular graphs of the following subsection. We are currently
investigating how to refine the treatment of non-regular graphs correspondingly.

For $d=3$, on the other hand, this improvement establishes the conjectured value
$\beta _3=16^{1/3}$, as is clear from the fact that $3f_{3,3}=16$ and $f_{2,3}>16^{1/3}$.

\subsubsection{Non-regular graphs}

We now consider other graphs, namely those with an initial cut of size less
than $d$.
As noted above this case is not the critical one and the argument is slightly more
messy and we only sketch the details.
For sufficiently small initial cuts, say $\le small_d$, this follows at once by induction
on the order of the graph  because $bound^{cut-1} > multiplier$
where $bound$ is the claimed bound on $\beta_d$, $cut$ is the cut size and
$multiplier$ is $f_{cut,d}$.
The values of $small_d$ for $d$ from $3$ to $11$ are
$[2,~3,~3,~4,~4,~5,~5,~6,~7]$.
For graphs with an initial cut of size between $small_d+1$ and $d-1$,
we modify the linear program and find that its solution
is greater than or equal to that obtained for the regular case.
First we improve the constraint
concerning $x_d$ to $x_d~=~ 0$ but we no longer have all the constraints
given by the average cut lemma but we do have them for $k\le small_d+1$.
We can moreover add new variables for the number of small cuts preceding the first
$i$-cut for $small_d+1 <  i<d$ and include the average cut lemma for the others.
Finally we can use the argument that, if up to some stage in the process (such as
the first such $i$-cut), the product of multipliers is sufficiently large,
the result follows by induction, so we can add to the linear program a constraint
that this does not happen.

In fact, for the application to memorisation mentioned in Section 1, we can
assume that the graph is not regular for reasons given in \cite{robson} but
the result for non-regular graphs is not of enough interest to merit
detailed study here.

\section{Conclusions}

For degree bounds up to $11$ we have shown that the number of spanning trees grows
at least exponentially with the cyclomatic number of a graph and we have shown
lower and upper bounds on the base of the exponent. The methods used are apparently
hard to generalise.

It would be much more satisfactory to have general proofs of any of the
three properties which we
have conjectured or proved for small $d$:
\begin{itemize}
\item $\beta_d$ is given by the complete graph $K_{d+1}$
\item Adding a new vertex of given degree to a graph $G$ multiplies $SP(G)$
by a factor which is minimised, over all graphs $G$ such that the resulting
graph has degree bounded by $d$, when $G$ is $K_{d}$
\item Cutting a graph $G$ (of degree bounded by $d$) into two parts $G_1$ and $G_2$
gives the minimum possible value of $SP(G)/(SP(G_1)SP(G_2))$ when $G_1$ or $G_2$ is
a single vertex.
\end{itemize}

\begin{table}[h]
\begin{tabular}{ccccccccc}
d=3 &4&5&6&7&8&9&10&11\\
$16^{1/3}$ & 2.236068 & 2.047672 & 1.912931 & 1.811447 & 1.732051 &
 1.668100 & 1.615394 & 1.571140\\
$16^{1/3}$ & 2.143571 & 1.959762 & 1.817549 & 1.725940 & 1.647326 &
 1.591588 & 1.541248 & 1.503335
\end{tabular}
\caption{Upper and lower bounds on $\beta_d$}
\end{table}


\bibliographystyle{splncs}

\end{document}


{\large\bf Appendix}\\The following conditions are tested
to check that the multipliers for cuts with both parts having more than one vertex
are subject to the same lower bounds as those where one part is a singleton.
In each case the format is the same. Writing $d$ for the degree bound,
$c$ for the cut value, $i$ and $p$ for the number of independent and dependent pairs:\\
$(c-maxu-maxv)+f[maxu,d]+f[maxv,d]+(2*i+p)(f[2,d]-2) \ge f[c,d]$\
followed by the numerical values of the two sides.

\begin{verbatim}
Case of a cut of value 4, with maxu= 2 and maxv= 2
There are 2 dependent pairs incident at u1 or v1  (1 at u1 and 1 at v1)
For u and v greedy partition is (2, 2);  
2 independent and 2 dependent pairs (excluding pairs incident at u1 or v1)
Degree bound 5:  0+f[5,2]+f[5,2]+(2*2+2)(f[5,2]-2) >= f[5,4] (7.200000>6.912000)
Degree bound 6:  0+f[6,2]+f[6,2]+(2*2+2)(f[6,2]-2) >= f[6,4] (6.666664>6.351852)
Degree bound 7:  0+f[7,2]+f[7,2]+(2*2+2)(f[7,2]-2) >= f[7,4] (6.285712>5.970845)
Degree bound 8:  0+f[8,2]+f[8,2]+(2*2+2)(f[8,2]-2) >= f[8,4] (6.000000>5.695313)
Degree bound 9:  0+f[9,2]+f[9,2]+(2*2+2)(f[9,2]-2) >= f[9,4] (5.777776>5.486968)
Degree bound 10: 0+f[10,2]+f[10,2]+(2*2+2)(f[10,2]-2) >= f[10,4] (5.600000>5.324000)
Degree bound 11: 0+f[11,2]+f[11,2]+(2*2+2)(f[11,2]-2) >= f[11,4] (5.454544>5.193088)

Case of a cut of value 5, with maxu= 2 and maxv= 2
There are 2 dependent pairs incident at u1 or v1  (1 at u1 and 1 at v1)
For u and v greedy partition is (2, 2, 1);  
6 independent and 2 dependent pairs (excluding pairs incident at u1 or v1)
Degree bound 6:  1+f[6,2]+f[6,2]+(2*6+2)(f[6,2]-2) >= f[6,5] (10.333328>9.263117)
Degree bound 7:  1+f[7,2]+f[7,2]+(2*6+2)(f[7,2]-2) >= f[7,5] (9.571424>8.529779)
Degree bound 8:  1+f[8,2]+f[8,2]+(2*6+2)(f[8,2]-2) >= f[8,5] (9.000000>8.009033)
Degree bound 9:  1+f[9,2]+f[9,2]+(2*6+2)(f[9,2]-2) >= f[9,5] (8.555552>7.620790)
Degree bound 10: 1+f[10,2]+f[10,2]+(2*6+2)(f[10,2]-2) >= f[10,5] (8.200000>7.320500)
Degree bound 11: 1+f[11,2]+f[11,2]+(2*6+2)(f[11,2]-2) >= f[11,5] (7.909088>7.081484)

Case of a cut of value 5, with maxu= 3 and maxv= 2
There are 4 dependent pairs incident at u1 or v1  (3 at u1 and 1 at v1)
For u greedy partition is (3, 2);   for v greedy partition is (2, 2, 1);  
4 independent and 2 dependent pairs (excluding pairs incident at u1 or v1)
Degree bound 6:  0+f[6,3]+f[6,2]+(2*4+2)(f[6,2]-2) >= f[6,5] (9.749996>9.263117)
Degree bound 7:  0+f[7,3]+f[7,2]+(2*4+2)(f[7,2]-2) >= f[7,5] (9.061221>8.529779)
Degree bound 8:  0+f[8,3]+f[8,2]+(2*4+2)(f[8,2]-2) >= f[8,5] (8.546875>8.009033)
Degree bound 9:  0+f[9,3]+f[9,2]+(2*4+2)(f[9,2]-2) >= f[9,5] (8.148146>7.620790)
Degree bound 10: 0+f[10,3]+f[10,2]+(2*4+2)(f[10,2]-2) >= f[10,5] (7.830000>7.320500)
Degree bound 11: 0+f[11,3]+f[11,2]+(2*4+2)(f[11,2]-2) >= f[11,5] (7.570246>7.081484)

Case of a cut of value 6, with maxu= 2 and maxv= 2
There are 2 dependent pairs incident at u1 or v1  (1 at u1 and 1 at v1)
For u and v greedy partition is (2, 2, 2);  
9 independent and 4 dependent pairs (excluding pairs incident at u1 or v1)
Degree bound 7:  2+f[7,2]+f[7,2]+(2*9+4)(f[7,2]-2) >= f[7,6] (12.857136>11.697983)
Degree bound 8:  2+f[8,2]+f[8,2]+(2*9+4)(f[8,2]-2) >= f[8,6] (12.000000>10.812195)
Degree bound 9:  2+f[9,2]+f[9,2]+(2*9+4)(f[9,2]-2) >= f[9,6] (11.333328>10.161053)
Degree bound 10: 2+f[10,2]+f[10,2]+(2*9+4)(f[10,2]-2) >= f[10,6] (10.800000>9.663060)
Degree bound 11: 2+f[11,2]+f[11,2]+(2*9+4)(f[11,2]-2) >= f[11,6] (10.363632>9.270306)

Case of a cut of value 6, with maxu= 3 and maxv= 2
There are 4 dependent pairs incident at u1 or v1  (3 at u1 and 1 at v1)
For u greedy partition is (3, 3);   for v greedy partition is (2, 2, 2);  
6 independent and 5 dependent pairs (excluding pairs incident at u1 or v1)
Degree bound 7:  1+f[7,3]+f[7,2]+(2*6+5)(f[7,2]-2) >= f[7,6] (12.061219>11.697983)
Degree bound 8:  1+f[8,3]+f[8,2]+(2*6+5)(f[8,2]-2) >= f[8,6] (11.296875>10.812195)
Degree bound 9:  1+f[9,3]+f[9,2]+(2*6+5)(f[9,2]-2) >= f[9,6] (10.703700>10.161053)
Degree bound 10: 1+f[10,3]+f[10,2]+(2*6+5)(f[10,2]-2) >= f[10,6] (10.230000>9.663060)
Degree bound 11: 1+f[11,3]+f[11,2]+(2*6+5)(f[11,2]-2) >= f[11,6] (9.842972>9.270306)

Case of a cut of value 6, with maxu= 3 and maxv= 3
There are 6 dependent pairs incident at u1 or v1  (3 at u1 and 3 at v1)
For u and v greedy partition is (3, 2, 1);  
7 independent and 2 dependent pairs (excluding pairs incident at u1 or v1)
Degree bound 7:  0+f[7,3]+f[7,3]+(2*7+2)(f[7,2]-2) >= f[7,6] (12.408158>11.697983)
Degree bound 8:  0+f[8,3]+f[8,3]+(2*7+2)(f[8,2]-2) >= f[8,6] (11.593750>10.812195)
Degree bound 9:  0+f[9,3]+f[9,3]+(2*7+2)(f[9,2]-2) >= f[9,6] (10.962960>10.161053)
Degree bound 10: 0+f[10,3]+f[10,3]+(2*7+2)(f[10,2]-2) >= f[10,6] (10.460000>9.663060)
Degree bound 11: 0+f[11,3]+f[11,3]+(2*7+2)(f[11,2]-2) >= f[11,6] (10.049584>9.270306)

Case of a cut of value 6, with maxu= 4 and maxv= 2
There are 7 dependent pairs incident at u1 or v1  (6 at u1 and 1 at v1)
For u greedy partition is (4, 2);   for v greedy partition is (2, 2, 1, 1);  
6 independent and 2 dependent pairs (excluding pairs incident at u1 or v1)
Degree bound 7:  0+f[7,4]+f[7,2]+(2*6+2)(f[7,2]-2) >= f[7,6] (12.256555>11.697983)
Degree bound 8:  0+f[8,4]+f[8,2]+(2*6+2)(f[8,2]-2) >= f[8,6] (11.445313>10.812195)
Degree bound 9:  0+f[9,4]+f[9,2]+(2*6+2)(f[9,2]-2) >= f[9,6] (10.820298>10.161053)
Degree bound 10: 0+f[10,4]+f[10,2]+(2*6+2)(f[10,2]-2) >= f[10,6] (10.324000>9.663060)
Degree bound 11: 0+f[11,4]+f[11,2]+(2*6+2)(f[11,2]-2) >= f[11,6] (9.920358>9.270306)

Case of a cut of value 7, with maxu= 2 and maxv= 2
There are 2 dependent pairs incident at u1 or v1  (1 at u1 and 1 at v1)
For u and v greedy partition is (2, 2, 2, 1);  
15 independent and 4 dependent pairs (excluding pairs incident at u1 or v1)
Degree bound 8:  3+f[8,2]+f[8,2]+(2*15+4)(f[8,2]-2) >= f[8,7] (16.000000>14.191006)
Degree bound 9:  3+f[9,2]+f[9,2]+(2*15+4)(f[9,2]-2) >= f[9,7] (14.999992>13.171735)
Degree bound 10: 3+f[10,2]+f[10,2]+(2*15+4)(f[10,2]-2) >= f[10,7] (14.200000>12.400927)
Degree bound 11: 3+f[11,2]+f[11,2]+(2*15+4)(f[11,2]-2) >= f[11,7] (13.545448>11.798571)

Case of a cut of value 7, with maxu= 3 and maxv= 2
There are 4 dependent pairs incident at u1 or v1  (3 at u1 and 1 at v1)
For u greedy partition is (3, 3, 1);   for v greedy partition is (2, 2, 2, 1);  
12 independent and 5 dependent pairs (excluding pairs incident at u1 or v1)
Degree bound 8:  2+f[8,3]+f[8,2]+(2*12+5)(f[8,2]-2) >= f[8,7] (15.296875>14.191006)
Degree bound 9:  2+f[9,3]+f[9,2]+(2*12+5)(f[9,2]-2) >= f[9,7] (14.370364>13.171735)
Degree bound 10: 2+f[10,3]+f[10,2]+(2*12+5)(f[10,2]-2) >= f[10,7] (13.630000>12.400927)
Degree bound 11: 2+f[11,3]+f[11,2]+(2*12+5)(f[11,2]-2) >= f[11,7] (13.024788>11.798571)

Case of a cut of value 7, with maxu= 3 and maxv= 3
There are 6 dependent pairs incident at u1 or v1  (3 at u1 and 3 at v1)
For u and v greedy partition is (3, 3, 1);  
9 independent and 6 dependent pairs (excluding pairs incident at u1 or v1)
Degree bound 8:  1+f[8,3]+f[8,3]+(2*9+6)(f[8,2]-2) >= f[8,7] (14.593750>14.191006)
Degree bound 9:  1+f[9,3]+f[9,3]+(2*9+6)(f[9,2]-2) >= f[9,7] (13.740736>13.171735)
Degree bound 10: 1+f[10,3]+f[10,3]+(2*9+6)(f[10,2]-2) >= f[10,7] (13.060000>12.400927)
Degree bound 11: 1+f[11,3]+f[11,3]+(2*9+6)(f[11,2]-2) >= f[11,7] (12.504128>11.798571)

Case of a cut of value 7, with maxu= 4 and maxv= 2
There are 7 dependent pairs incident at u1 or v1  (6 at u1 and 1 at v1)
For u greedy partition is (4, 3);   for v greedy partition is (2, 2, 2, 1);  
9 independent and 5 dependent pairs (excluding pairs incident at u1 or v1)
Degree bound 8:  1+f[8,4]+f[8,2]+(2*9+5)(f[8,2]-2) >= f[8,7] (14.695313>14.191006)
Degree bound 9:  1+f[9,4]+f[9,2]+(2*9+5)(f[9,2]-2) >= f[9,7] (13.820296>13.171735)
Degree bound 10: 1+f[10,4]+f[10,2]+(2*9+5)(f[10,2]-2) >= f[10,7] (13.124000>12.400927)
Degree bound 11: 1+f[11,4]+f[11,2]+(2*9+5)(f[11,2]-2) >= f[11,7] (12.556720>11.798571)

Case of a cut of value 7, with maxu= 4 and maxv= 3
There are 9 dependent pairs incident at u1 or v1  (6 at u1 and 3 at v1)
For u greedy partition is (4, 2, 1);   for v greedy partition is (3, 2, 1, 1);  
10 independent and 2 dependent pairs (excluding pairs incident at u1 or v1)
Degree bound 8:  0+f[8,4]+f[8,3]+(2*10+2)(f[8,2]-2) >= f[8,7] (14.992188>14.191006)
Degree bound 9:  0+f[9,4]+f[9,3]+(2*10+2)(f[9,2]-2) >= f[9,7] (14.079556>13.171735)
Degree bound 10: 0+f[10,4]+f[10,3]+(2*10+2)(f[10,2]-2) >= f[10,7] (13.354000>12.400927)
Degree bound 11: 0+f[11,4]+f[11,3]+(2*10+2)(f[11,2]-2) >= f[11,7] (12.763332>11.798571)

Case of a cut of value 7, with maxu= 5 and maxv= 2
There are 11 dependent pairs incident at u1 or v1  (10 at u1 and 1 at v1)
For u greedy partition is (5, 2);   for v greedy partition is (2, 2, 1, 1, 1);  
8 independent and 2 dependent pairs (excluding pairs incident at u1 or v1)
Degree bound 8:  0+f[8,5]+f[8,2]+(2*8+2)(f[8,2]-2) >= f[8,7] (14.759033>14.191006)
Degree bound 9:  0+f[9,5]+f[9,2]+(2*8+2)(f[9,2]-2) >= f[9,7] (13.843008>13.171735)
Degree bound 10: 0+f[10,5]+f[10,2]+(2*8+2)(f[10,2]-2) >= f[10,7] (13.120500>12.400927)
Degree bound 11: 0+f[11,5]+f[11,2]+(2*8+2)(f[11,2]-2) >= f[11,7] (12.536026>11.798571)

Case of a cut of value 8, with maxu= 2 and maxv= 2
There are 2 dependent pairs incident at u1 or v1  (1 at u1 and 1 at v1)
For u and v greedy partition is (2, 2, 2, 2);  
20 independent and 6 dependent pairs (excluding pairs incident at u1 or v1)
Degree bound 9:  4+f[9,2]+f[9,2]+(2*20+6)(f[9,2]-2) >= f[9,8] (18.666656>16.726013)
Degree bound 10: 4+f[10,2]+f[10,2]+(2*20+6)(f[10,2]-2) >= f[10,8] (17.600000>15.589737)
Degree bound 11: 4+f[11,2]+f[11,2]+(2*20+6)(f[11,2]-2) >= f[11,8] (16.727264>14.709907)

Case of a cut of value 8, with maxu= 3 and maxv= 2
There are 4 dependent pairs incident at u1 or v1  (3 at u1 and 1 at v1)
For u greedy partition is (3, 3, 2);   for v greedy partition is (2, 2, 2, 2);  
17 independent and 7 dependent pairs (excluding pairs incident at u1 or v1)
Degree bound 9:  3+f[9,3]+f[9,2]+(2*17+7)(f[9,2]-2) >= f[9,8] (18.037028>16.726013)
Degree bound 10: 3+f[10,3]+f[10,2]+(2*17+7)(f[10,2]-2) >= f[10,8] (17.030000>15.589737)
Degree bound 11: 3+f[11,3]+f[11,2]+(2*17+7)(f[11,2]-2) >= f[11,8] (16.206604>14.709907)

Case of a cut of value 8, with maxu= 3 and maxv= 3
There are 6 dependent pairs incident at u1 or v1  (3 at u1 and 3 at v1)
For u and v greedy partition is (3, 3, 2);  
14 independent and 8 dependent pairs (excluding pairs incident at u1 or v1)
Degree bound 9:  2+f[9,3]+f[9,3]+(2*14+8)(f[9,2]-2) >= f[9,8] (17.407400>16.726013)
Degree bound 10: 2+f[10,3]+f[10,3]+(2*14+8)(f[10,2]-2) >= f[10,8] (16.460000>15.589737)
Degree bound 11: 2+f[11,3]+f[11,3]+(2*14+8)(f[11,2]-2) >= f[11,8] (15.685944>14.709907)

Case of a cut of value 8, with maxu= 4 and maxv= 2
There are 7 dependent pairs incident at u1 or v1  (6 at u1 and 1 at v1)
For u greedy partition is (4, 4);   for v greedy partition is (2, 2, 2, 2);  
12 independent and 9 dependent pairs (excluding pairs incident at u1 or v1)
Degree bound 9:  2+f[9,4]+f[9,2]+(2*12+9)(f[9,2]-2) >= f[9,8] (17.042516>16.726013)
Degree bound 10: 2+f[10,4]+f[10,2]+(2*12+9)(f[10,2]-2) >= f[10,8] (16.124000>15.589737)
Degree bound 11: 2+f[11,4]+f[11,2]+(2*12+9)(f[11,2]-2) >= f[11,8] (15.374900>14.709907)

Case of a cut of value 8, with maxu= 4 and maxv= 3
There are 9 dependent pairs incident at u1 or v1  (6 at u1 and 3 at v1)
For u greedy partition is (4, 3, 1);   for v greedy partition is (3, 3, 1, 1);  
13 independent and 6 dependent pairs (excluding pairs incident at u1 or v1)
Degree bound 9:  1+f[9,4]+f[9,3]+(2*13+6)(f[9,2]-2) >= f[9,8] (17.301776>16.726013)
Degree bound 10: 1+f[10,4]+f[10,3]+(2*13+6)(f[10,2]-2) >= f[10,8] (16.354000>15.589737)
Degree bound 11: 1+f[11,4]+f[11,3]+(2*13+6)(f[11,2]-2) >= f[11,8] (15.581512>14.709907)

Case of a cut of value 8, with maxu= 4 and maxv= 4
There are 12 dependent pairs incident at u1 or v1  (6 at u1 and 6 at v1)
For u and v greedy partition is (4, 2, 1, 1);  
14 independent and 2 dependent pairs (excluding pairs incident at u1 or v1)
Degree bound 9:  0+f[9,4]+f[9,4]+(2*14+2)(f[9,2]-2) >= f[9,8] (17.640596>16.726013)
Degree bound 10: 0+f[10,4]+f[10,4]+(2*14+2)(f[10,2]-2) >= f[10,8] (16.648000>15.589737)
Degree bound 11: 0+f[11,4]+f[11,4]+(2*14+2)(f[11,2]-2) >= f[11,8] (15.840716>14.709907)

Case of a cut of value 8, with maxu= 5 and maxv= 2
There are 11 dependent pairs incident at u1 or v1  (10 at u1 and 1 at v1)
For u greedy partition is (5, 3);   for v greedy partition is (2, 2, 2, 1, 1);  
12 independent and 5 dependent pairs (excluding pairs incident at u1 or v1)
Degree bound 9:  1+f[9,5]+f[9,2]+(2*12+5)(f[9,2]-2) >= f[9,8] (17.287450>16.726013)
Degree bound 10: 1+f[10,5]+f[10,2]+(2*12+5)(f[10,2]-2) >= f[10,8] (16.320500>15.589737)
Degree bound 11: 1+f[11,5]+f[11,2]+(2*12+5)(f[11,2]-2) >= f[11,8] (15.536024>14.709907)

Case of a cut of value 8, with maxu= 5 and maxv= 3
There are 13 dependent pairs incident at u1 or v1  (10 at u1 and 3 at v1)
For u greedy partition is (5, 2, 1);   for v greedy partition is (3, 2, 1, 1, 1);  
13 independent and 2 dependent pairs (excluding pairs incident at u1 or v1)
Degree bound 9:  0+f[9,5]+f[9,3]+(2*13+2)(f[9,2]-2) >= f[9,8] (17.546710>16.726013)
Degree bound 10: 0+f[10,5]+f[10,3]+(2*13+2)(f[10,2]-2) >= f[10,8] (16.550500>15.589737)
Degree bound 11: 0+f[11,5]+f[11,3]+(2*13+2)(f[11,2]-2) >= f[11,8] (15.742636>14.709907)

Case of a cut of value 8, with maxu= 6 and maxv= 2
There are 16 dependent pairs incident at u1 or v1  (15 at u1 and 1 at v1)
For u greedy partition is (6, 2);   for v greedy partition is (2, 2, 1, 1, 1, 1);  
10 independent and 2 dependent pairs (excluding pairs incident at u1 or v1)
Degree bound 9:  0+f[9,6]+f[9,2]+(2*10+2)(f[9,2]-2) >= f[9,8] (17.272159>16.726013)
Degree bound 10: 0+f[10,6]+f[10,2]+(2*10+2)(f[10,2]-2) >= f[10,8] (16.263060>15.589737)
Degree bound 11: 0+f[11,6]+f[11,2]+(2*10+2)(f[11,2]-2) >= f[11,8] (15.452120>14.709907)

Case of a cut of value 9, with maxu= 2 and maxv= 2
There are 2 dependent pairs incident at u1 or v1  (1 at u1 and 1 at v1)
For u and v greedy partition is (2, 2, 2, 2, 1);  
28 independent and 6 dependent pairs (excluding pairs incident at u1 or v1)
Degree bound 10: 5+f[10,2]+f[10,2]+(2*28+6)(f[10,2]-2) >= f[10,9] (21.800000>19.292299)
Degree bound 11: 5+f[11,2]+f[11,2]+(2*28+6)(f[11,2]-2) >= f[11,9] (20.636352>18.053067)

Case of a cut of value 9, with maxu= 3 and maxv= 2
There are 4 dependent pairs incident at u1 or v1  (3 at u1 and 1 at v1)
For u greedy partition is (3, 3, 3);   for v greedy partition is (2, 2, 2, 2, 1);  
23 independent and 9 dependent pairs (excluding pairs incident at u1 or v1)
Degree bound 10: 4+f[10,3]+f[10,2]+(2*23+9)(f[10,2]-2) >= f[10,9] (20.830000>19.292299)
Degree bound 11: 4+f[11,3]+f[11,2]+(2*23+9)(f[11,2]-2) >= f[11,9] (19.752056>18.053067)

Case of a cut of value 9, with maxu= 3 and maxv= 3
There are 6 dependent pairs incident at u1 or v1  (3 at u1 and 3 at v1)
For u and v greedy partition is (3, 3, 3);  
18 independent and 12 dependent pairs (excluding pairs incident at u1 or v1)
Degree bound 10: 3+f[10,3]+f[10,3]+(2*18+12)(f[10,2]-2) >= f[10,9] (19.860000>19.292299)
Degree bound 11: 3+f[11,3]+f[11,3]+(2*18+12)(f[11,2]-2) >= f[11,9] (18.867760>18.053067)

Case of a cut of value 9, with maxu= 4 and maxv= 2
There are 7 dependent pairs incident at u1 or v1  (6 at u1 and 1 at v1)
For u greedy partition is (4, 4, 1);   for v greedy partition is (2, 2, 2, 2, 1);  
20 independent and 9 dependent pairs (excluding pairs incident at u1 or v1)
Degree bound 10: 3+f[10,4]+f[10,2]+(2*20+9)(f[10,2]-2) >= f[10,9] (20.324000>19.292299)
Degree bound 11: 3+f[11,4]+f[11,2]+(2*20+9)(f[11,2]-2) >= f[11,9] (19.283988>18.053067)

Case of a cut of value 9, with maxu= 4 and maxv= 3
There are 9 dependent pairs incident at u1 or v1  (6 at u1 and 3 at v1)
For u greedy partition is (4, 4, 1);   for v greedy partition is (3, 3, 2, 1);  
17 independent and 10 dependent pairs (excluding pairs incident at u1 or v1)
Degree bound 10: 2+f[10,4]+f[10,3]+(2*17+10)(f[10,2]-2) >= f[10,9] (19.754000>19.292299)
Degree bound 11: 2+f[11,4]+f[11,3]+(2*17+10)(f[11,2]-2) >= f[11,9] (18.763328>18.053067)

Case of a cut of value 9, with maxu= 4 and maxv= 4
There are 12 dependent pairs incident at u1 or v1  (6 at u1 and 6 at v1)
For u and v greedy partition is (4, 3, 1, 1);  
18 independent and 6 dependent pairs (excluding pairs incident at u1 or v1)
Degree bound 10: 1+f[10,4]+f[10,4]+(2*18+6)(f[10,2]-2) >= f[10,9] (20.048000>19.292299)
Degree bound 11: 1+f[11,4]+f[11,4]+(2*18+6)(f[11,2]-2) >= f[11,9] (19.022532>18.053067)

Case of a cut of value 9, with maxu= 5 and maxv= 2
There are 11 dependent pairs incident at u1 or v1  (10 at u1 and 1 at v1)
For u greedy partition is (5, 4);   for v greedy partition is (2, 2, 2, 2, 1);  
16 independent and 9 dependent pairs (excluding pairs incident at u1 or v1)
Degree bound 10: 2+f[10,5]+f[10,2]+(2*16+9)(f[10,2]-2) >= f[10,9] (19.720500>19.292299)
Degree bound 11: 2+f[11,5]+f[11,2]+(2*16+9)(f[11,2]-2) >= f[11,9] (18.717840>18.053067)

Case of a cut of value 9, with maxu= 5 and maxv= 3
There are 13 dependent pairs incident at u1 or v1  (10 at u1 and 3 at v1)
For u greedy partition is (5, 3, 1);   for v greedy partition is (3, 3, 1, 1, 1);  
17 independent and 6 dependent pairs (excluding pairs incident at u1 or v1)
Degree bound 10: 1+f[10,5]+f[10,3]+(2*17+6)(f[10,2]-2) >= f[10,9] (19.950500>19.292299)
Degree bound 11: 1+f[11,5]+f[11,3]+(2*17+6)(f[11,2]-2) >= f[11,9] (18.924452>18.053067)

Case of a cut of value 9, with maxu= 5 and maxv= 4
There are 16 dependent pairs incident at u1 or v1  (10 at u1 and 6 at v1)
For u greedy partition is (5, 2, 1, 1);   for v greedy partition is (4, 2, 1, 1, 1);  
18 independent and 2 dependent pairs (excluding pairs incident at u1 or v1)
Degree bound 10: 0+f[10,5]+f[10,4]+(2*18+2)(f[10,2]-2) >= f[10,9] (20.244500>19.292299)
Degree bound 11: 0+f[11,5]+f[11,4]+(2*18+2)(f[11,2]-2) >= f[11,9] (19.183656>18.053067)

Case of a cut of value 9, with maxu= 6 and maxv= 2
There are 16 dependent pairs incident at u1 or v1  (15 at u1 and 1 at v1)
For u greedy partition is (6, 3);   for v greedy partition is (2, 2, 2, 1, 1, 1);  
15 independent and 5 dependent pairs (excluding pairs incident at u1 or v1)
Degree bound 10: 1+f[10,6]+f[10,2]+(2*15+5)(f[10,2]-2) >= f[10,9] (19.863060>19.292299)
Degree bound 11: 1+f[11,6]+f[11,2]+(2*15+5)(f[11,2]-2) >= f[11,9] (18.815754>18.053067)

Case of a cut of value 9, with maxu= 6 and maxv= 3
There are 18 dependent pairs incident at u1 or v1  (15 at u1 and 3 at v1)
For u greedy partition is (6, 2, 1);   for v greedy partition is (3, 2, 1, 1, 1, 1);  
16 independent and 2 dependent pairs (excluding pairs incident at u1 or v1)
Degree bound 10: 0+f[10,6]+f[10,3]+(2*16+2)(f[10,2]-2) >= f[10,9] (20.093060>19.292299)
Degree bound 11: 0+f[11,6]+f[11,3]+(2*16+2)(f[11,2]-2) >= f[11,9] (19.022366>18.053067)

Case of a cut of value 9, with maxu= 7 and maxv= 2
There are 22 dependent pairs incident at u1 or v1  (21 at u1 and 1 at v1)
For u greedy partition is (7, 2);   for v greedy partition is (2, 2, 1, 1, 1, 1, 1);  
12 independent and 2 dependent pairs (excluding pairs incident at u1 or v1)
Degree bound 10: 0+f[10,7]+f[10,2]+(2*12+2)(f[10,2]-2) >= f[10,9] (19.800927>19.292299)
Degree bound 11: 0+f[11,7]+f[11,2]+(2*12+2)(f[11,2]-2) >= f[11,9] (18.707657>18.053067)

Case of a cut of value 10, with maxu= 2 and maxv= 2
There are 2 dependent pairs incident at u1 or v1  (1 at u1 and 1 at v1)
For u and v greedy partition is (2, 2, 2, 2, 2);  
35 independent and 8 dependent pairs (excluding pairs incident at u1 or v1)
Degree bound 11: 6+f[11,2]+f[11,2]+(2*35+8)(f[11,2]-2) >= f[11,10] (24.545440>21.882506)

Case of a cut of value 10, with maxu= 3 and maxv= 2
There are 4 dependent pairs incident at u1 or v1  (3 at u1 and 1 at v1)
For u greedy partition is (3, 3, 3, 1);   for v greedy partition is (2, 2, 2, 2, 2);  
31 independent and 10 dependent pairs (excluding pairs incident at u1 or v1)
Degree bound 11: 5+f[11,3]+f[11,2]+(2*31+10)(f[11,2]-2) >= f[11,10] (23.842962>21.882506)

Case of a cut of value 10, with maxu= 3 and maxv= 3
There are 6 dependent pairs incident at u1 or v1  (3 at u1 and 3 at v1)
For u and v greedy partition is (3, 3, 3, 1);  
27 independent and 12 dependent pairs (excluding pairs incident at u1 or v1)
Degree bound 11: 4+f[11,3]+f[11,3]+(2*27+12)(f[11,2]-2) >= f[11,10] (23.140484>21.882506)

Case of a cut of value 10, with maxu= 4 and maxv= 2
There are 7 dependent pairs incident at u1 or v1  (6 at u1 and 1 at v1)
For u greedy partition is (4, 4, 2);   for v greedy partition is (2, 2, 2, 2, 2);  
27 independent and 11 dependent pairs (excluding pairs incident at u1 or v1)
Degree bound 11: 4+f[11,4]+f[11,2]+(2*27+11)(f[11,2]-2) >= f[11,10] (23.193076>21.882506)

Case of a cut of value 10, with maxu= 4 and maxv= 3
There are 9 dependent pairs incident at u1 or v1  (6 at u1 and 3 at v1)
For u greedy partition is (4, 4, 2);   for v greedy partition is (3, 3, 3, 1);  
23 independent and 13 dependent pairs (excluding pairs incident at u1 or v1)
Degree bound 11: 3+f[11,4]+f[11,3]+(2*23+13)(f[11,2]-2) >= f[11,10] (22.490598>21.882506)

Case of a cut of value 10, with maxu= 4 and maxv= 4
There are 12 dependent pairs incident at u1 or v1  (6 at u1 and 6 at v1)
For u and v greedy partition is (4, 4, 1, 1);  
21 independent and 12 dependent pairs (excluding pairs incident at u1 or v1)
Degree bound 11: 2+f[11,4]+f[11,4]+(2*21+12)(f[11,2]-2) >= f[11,10] (22.204348>21.882506)

Case of a cut of value 10, with maxu= 5 and maxv= 2
There are 11 dependent pairs incident at u1 or v1  (10 at u1 and 1 at v1)
For u greedy partition is (5, 5);   for v greedy partition is (2, 2, 2, 2, 2);  
20 independent and 14 dependent pairs (excluding pairs incident at u1 or v1)
Degree bound 11: 3+f[11,5]+f[11,2]+(2*20+14)(f[11,2]-2) >= f[11,10] (22.081474>21.882506)

Case of a cut of value 10, with maxu= 5 and maxv= 3
There are 13 dependent pairs incident at u1 or v1  (10 at u1 and 3 at v1)
For u greedy partition is (5, 4, 1);   for v greedy partition is (3, 3, 2, 1, 1);  
22 independent and 10 dependent pairs (excluding pairs incident at u1 or v1)
Degree bound 11: 2+f[11,5]+f[11,3]+(2*22+10)(f[11,2]-2) >= f[11,10] (22.469904>21.882506)

Case of a cut of value 10, with maxu= 5 and maxv= 4
There are 16 dependent pairs incident at u1 or v1  (10 at u1 and 6 at v1)
For u greedy partition is (5, 3, 1, 1);   for v greedy partition is (4, 3, 1, 1, 1);  
23 independent and 6 dependent pairs (excluding pairs incident at u1 or v1)
Degree bound 11: 1+f[11,5]+f[11,4]+(2*23+6)(f[11,2]-2) >= f[11,10] (22.729108>21.882506)

Case of a cut of value 10, with maxu= 5 and maxv= 5
There are 20 dependent pairs incident at u1 or v1  (10 at u1 and 10 at v1)
For u and v greedy partition is (5, 2, 1, 1, 1);  
23 independent and 2 dependent pairs (excluding pairs incident at u1 or v1)
Degree bound 11: 0+f[11,5]+f[11,5]+(2*23+2)(f[11,2]-2) >= f[11,10] (22.890232>21.882506)

Case of a cut of value 10, with maxu= 6 and maxv= 2
There are 16 dependent pairs incident at u1 or v1  (15 at u1 and 1 at v1)
For u greedy partition is (6, 4);   for v greedy partition is (2, 2, 2, 2, 1, 1);  
20 independent and 9 dependent pairs (excluding pairs incident at u1 or v1)
Degree bound 11: 2+f[11,6]+f[11,2]+(2*20+9)(f[11,2]-2) >= f[11,10] (22.361206>21.882506)

Case of a cut of value 10, with maxu= 6 and maxv= 3
There are 18 dependent pairs incident at u1 or v1  (15 at u1 and 3 at v1)
For u greedy partition is (6, 3, 1);   for v greedy partition is (3, 3, 1, 1, 1, 1);  
21 independent and 6 dependent pairs (excluding pairs incident at u1 or v1)
Degree bound 11: 1+f[11,6]+f[11,3]+(2*21+6)(f[11,2]-2) >= f[11,10] (22.567818>21.882506)

Case of a cut of value 10, with maxu= 6 and maxv= 4
There are 21 dependent pairs incident at u1 or v1  (15 at u1 and 6 at v1)
For u greedy partition is (6, 2, 1, 1);   for v greedy partition is (4, 2, 1, 1, 1, 1);  
22 independent and 2 dependent pairs (excluding pairs incident at u1 or v1)
Degree bound 11: 0+f[11,6]+f[11,4]+(2*22+2)(f[11,2]-2) >= f[11,10] (22.827022>21.882506)

Case of a cut of value 10, with maxu= 7 and maxv= 2
There are 22 dependent pairs incident at u1 or v1  (21 at u1 and 1 at v1)
For u greedy partition is (7, 3);   for v greedy partition is (2, 2, 2, 1, 1, 1, 1);  
18 independent and 5 dependent pairs (excluding pairs incident at u1 or v1)
Degree bound 11: 1+f[11,7]+f[11,2]+(2*18+5)(f[11,2]-2) >= f[11,10] (22.434927>21.882506)

Case of a cut of value 10, with maxu= 7 and maxv= 3
There are 24 dependent pairs incident at u1 or v1  (21 at u1 and 3 at v1)
For u greedy partition is (7, 2, 1);   for v greedy partition is (3, 2, 1, 1, 1, 1, 1);  
19 independent and 2 dependent pairs (excluding pairs incident at u1 or v1)
Degree bound 11: 0+f[11,7]+f[11,3]+(2*19+2)(f[11,2]-2) >= f[11,10] (22.641539>21.882506)

Case of a cut of value 10, with maxu= 8 and maxv= 2
There are 29 dependent pairs incident at u1 or v1  (28 at u1 and 1 at v1)
For u greedy partition is (8, 2);   for v greedy partition is (2, 2, 1, 1, 1, 1, 1, 1);  
14 independent and 2 dependent pairs (excluding pairs incident at u1 or v1)
Degree bound 11: 0+f[11,8]+f[11,2]+(2*14+2)(f[11,2]-2) >= f[11,10] (22.346265>21.882506)

\end{verbatim}